\documentclass[twocolumn]{aa}

\usepackage{graphicx}
\usepackage{amsmath,amsfonts,amssymb}
\usepackage{txfonts}
\usepackage{color}
\usepackage{natbib}
\usepackage{float}
\usepackage{dblfloatfix}
\usepackage{afterpage}
\usepackage{ifthen}
\usepackage[morefloats=12]{morefloats}
\usepackage{placeins}
\usepackage{multicol}
\bibpunct{(}{)}{;}{a}{}{,}
\usepackage[switch]{lineno}
\definecolor{linkcolor}{rgb}{0.6,0,0}
\definecolor{citecolor}{rgb}{0,0,0.75}
\definecolor{urlcolor}{rgb}{0.12,0.46,0.7}
\usepackage[breaklinks, colorlinks, urlcolor=urlcolor,
    linkcolor=linkcolor,citecolor=citecolor,pdfencoding=auto]{hyperref}
\hypersetup{linktocpage}
\usepackage{bold-extra}
\usepackage{xcolor}

\def\setsymbol#1#2{\expandafter\def\csname #1\endcsname{#2}}
\def\getsymbol#1{\csname #1\endcsname}

\def\Planck{\textit{Planck}}

\newbox\tablebox    \newdimen\tablewidth
\def\leaderfil{\leaders\hbox to 5pt{\hss.\hss}\hfil}
\def\tablenote#1 #2\par{\begingroup \parindent=0.8em
    \abovedisplayshortskip=0pt\belowdisplayshortskip=0pt
    \noindent
    $$\hss\vbox{\hsize\tablewidth \hangindent=\parindent \hangafter=1 \noindent
    \hbox to \parindent{$^#1$\hss}\strut#2\strut\par}\hss$$
    \endgroup}

\def\L2{\ifmmode L_2\else $L_2$\fi}

\def\DeltaT{\ifmmode \Delta T\else $\Delta T$\fi}
\def\deltat{\ifmmode \Delta t\else $\Delta t$\fi}
\def\fknee{\ifmmode f_{\rm knee}\else $f_{\rm knee}$\fi}
\def\Fmax{\ifmmode F_{\rm max}\else $F_{\rm max}$\fi}
\def\solar{\ifmmode{\rm M}_{\mathord\odot}\else${\rm M}_{\mathord\odot}$\fi}
\def\Msolar{\ifmmode{\rm M}_{\mathord\odot}\else${\rm M}_{\mathord\odot}$\fi}
\def\Lsolar{\ifmmode{\rm L}_{\mathord\odot}\else${\rm L}_{\mathord\odot}$\fi}
\def\inv{\ifmmode^{-1}\else$^{-1}$\fi}
\def\mo{\ifmmode^{-1}\else$^{-1}$\fi}
\def\sup#1{\ifmmode ^{\rm #1}\else $^{\rm #1}$\fi}
\def\expo#1{\ifmmode \times 10^{#1}\else $\times 10^{#1}$\fi}
\def\,{\thinspace}
\def\lsim{\mathrel{\raise .4ex\hbox{\rlap{$<$}\lower 1.2ex\hbox{$\sim$}}}}
\def\gsim{\mathrel{\raise .4ex\hbox{\rlap{$>$}\lower 1.2ex\hbox{$\sim$}}}}

\def\simprop{\mathrel{\raise .4ex\hbox{\rlap{$\propto$}\lower 1.2ex\hbox{$\sim$}}}}
\def\deg{\ifmmode^\circ\else$^\circ$\fi}
\def\pdeg{\ifmmode $\setbox0=\hbox{$^{\circ}$}\rlap{\hskip.11\wd0 .}$^{\circ}
          \else \setbox0=\hbox{$^{\circ}$}\rlap{\hskip.11\wd0 .}$^{\circ}$\fi}
\def\arcs{\ifmmode {^{\scriptstyle\prime\prime}}
          \else $^{\scriptstyle\prime\prime}$\fi}
\def\arcm{\ifmmode {^{\scriptstyle\prime}}
          \else $^{\scriptstyle\prime}$\fi}
\newdimen\sa  \newdimen\sb
\def\parcs{\sa=.07em \sb=.03em
     \ifmmode \hbox{\rlap{.}}^{\scriptstyle\prime\kern -\sb\prime}\hbox{\kern -\sa}
     \else \rlap{.}$^{\scriptstyle\prime\kern -\sb\prime}$\kern -\sa\fi}
\def\parcm{\sa=.08em \sb=.03em
     \ifmmode \hbox{\rlap{.}\kern\sa}^{\scriptstyle\prime}\hbox{\kern-\sb}
     \else \rlap{.}\kern\sa$^{\scriptstyle\prime}$\kern-\sb\fi}
\def\ra[#1 #2 #3.#4]{#1\sup{h}#2\sup{m}#3\sup{s}\llap.#4}
\def\dec[#1 #2 #3.#4]{#1\deg#2\arcm#3\arcs\llap.#4}
\def\deco[#1 #2 #3]{#1\deg#2\arcm#3\arcs}
\def\rra[#1 #2]{#1\sup{h}#2\sup{m}}

\def\dots{\relax\ifmmode \ldots\else $\ldots$\fi}
\def\WHzsr{\ifmmode $W\,Hz\mo\,sr\mo$\else W\,Hz\mo\,sr\mo\fi}
\def\mHz{\ifmmode $\,mHz$\else \,mHz\fi}
\def\GHz{\ifmmode $\,GHz$\else \,GHz\fi}
\def\mKs{\ifmmode $\,mK\,s$^{1/2}\else \,mK\,s$^{1/2}$\fi}
\def\muKs{\ifmmode \,\mu$K\,s$^{1/2}\else \,$\mu$K\,s$^{1/2}$\fi}
\def\muKRJs{\ifmmode \,\mu$K$_{\rm RJ}$\,s$^{1/2}\else \,$\mu$K$_{\rm RJ}$\,s$^{1/2}$\fi}
\def\muKHz{\ifmmode \,\mu$K\,Hz$^{-1/2}\else \,$\mu$K\,Hz$^{-1/2}$\fi}
\def\MJysr{\ifmmode \,$MJy\,sr\mo$\else \,MJy\,sr\mo\fi}
\def\MJysrmK{\ifmmode \,$MJy\,sr\mo$\,mK$_{\rm CMB}\mo\else \,MJy\,sr\mo\,mK$_{\rm CMB}\mo$\fi}
\def\microns{\ifmmode \,\mu$m$\else \,$\mu$m\fi}

\def\muK{\ifmmode \,\mu$K$\else \,$\mu$\hbox{K}\fi}
\def\microK{\ifmmode \,\mu$K$\else \,$\mu$\hbox{K}\fi}
\def\muW{\ifmmode \,\mu$W$\else \,$\mu$\hbox{W}\fi}
\def\kms{\ifmmode $\,km\,s$^{-1}\else \,km\,s$^{-1}$\fi}
\def\kmsMpc{\ifmmode $\,\kms\,Mpc\mo$\else \,\kms\,Mpc\mo\fi}
\providecommand{\sorthelp}[1]{}

\def\planck{\emph{Planck}}
\def\Planck{\emph{Planck}}

\def\commander{\texttt{Commander}}

\def\commanderthree{\texttt{Commander3}}

\def\sroll2{\texttt{SRoll2}}
\def\cosmomc{\texttt{CosmoMC}}
\def\cobaya{\texttt{Cobaya}}
\def\camb{\texttt{CAMB}}

\renewcommand{\d}[0]{\vec{d}}

\newcommand{\B}[0]{\tens{B}}

\newcommand{\n}[0]{\vec{n}}

\newcommand{\s}[0]{\vec{s}}
\renewcommand{\a}[0]{\vec{a}}

\newcommand{\f}[0]{\vec{f}}

\renewcommand{\L}[0]{\tens{L}}

\newcommand{\N}[0]{\tens{N}}
\newcommand{\M}[0]{\tens{M}}

\newcommand{\w}[0]{\vec{w}}
\renewcommand{\S}[0]{\tens{S}}
\renewcommand{\r}[0]{\vec{r}}

\renewcommand{\P}[0]{\tens{P}}

\newcommand{\BP}{\textsc{BeyondPlanck}}

\newcommand{\cosmoglobe}{\textsc{Cosmoglobe}}
\newcommand{\Cosmoglobe}{\textsc{Cosmoglobe}}

\newcommand{\data}{\vec d}
\newcommand{\ncorr}{\vec n_\mathrm{corr}}

\def\inv{^{-1}}

\begin{document}

\title{\bfseries{\Cosmoglobe: Towards end-to-end CMB cosmological\\ parameter estimation without likelihood approximations}}
\newcommand{\oslo}[0]{1}
\newcommand{\iiabangalore}[0]{2}

\author{\small
J.~R.~Eskilt\inst{\ref{uio},\ref{imperial}}\thanks{Corresponding author: J.~R.~Eskilt; \url{j.r.eskilt@astro.uio.no}}
\and
K.~Lee\inst{\ref{uio}}
\and
D.~J.~Watts\inst{\ref{uio}}
\and
V.~Anshul\inst{\ref{iit_bhu}}
\and
R.~Aurlien\inst{\ref{uio}}
\and
A.~Basyrov\inst{\ref{uio}}
\and
M.~Bersanelli\inst{\ref{milan}}
\and
L.~P.~L.~Colombo\inst{\ref{milan}}
\and
H.~K.~Eriksen\inst{\ref{uio}}
\and
K.~S.~F.~Fornazier\inst{\ref{saopaulo}}
\and
U.~Fuskeland\inst{\ref{uio}}
\and
M.~Galloway\inst{\ref{uio}}
\and
E.~Gjerl\o w\inst{\ref{uio}}
\and
L.~T.~Hergt\inst{\ref{ubc}}
%\and
%G.~A.~Hoerning\inst{\ref{saopaulo}}
\and
H.~T.~Ihle\inst{\ref{uio}}
\and
J.~G.~S.~Lunde\inst{\ref{uio}}
\and
A.~Marins\inst{\ref{saopaulo}}
\and
S.~K.~Nerval\inst{\ref{dunlap1},\ref{dunlap2}}
\and
S.~Paradiso\inst{\ref{waterloo},\ref{waterloo2}}
\and
F.~Rahman\inst{\ref{iiabangalore}}
\and
M.~San\inst{\ref{uio}}
\and
N.-O.~Stutzer\inst{\ref{uio}}
\and
I.~K.~Wehus\inst{\ref{uio}}
}
\institute{\small
Institute of Theoretical Astrophysics, University of Oslo, Blindern, Oslo, Norway\label{uio}
\and
Imperial Centre for Inference and Cosmology, Department of Physics, Imperial College London, Blackett Laboratory, Prince Consort Road, London SW7 2AZ, United Kingdom\label{imperial}
\and
Department of Physics, Indian Institute of Technology (BHU), Varanasi - 221005, India\label{iit_bhu}
\and
Dipartimento di Fisica, Università degli Studi di Milano, Via Celoria, 16, Milano, Italy\label{milan}
\and
Instituto de Física, Universidade de São Paulo - C.P. 66318, CEP: 05315-970, São Paulo, Brazil\label{saopaulo}
\and
Department of Physics and Astronomy, University of British Columbia, 6224 Agricultural Road, Vancouver BC, V6T1Z1, Canada\label{ubc}
\and
David A. Dunlap Department of Astronomy \& Astrophysics, University of Toronto, 50 St. George Street, Toronto, ON M5S 3H4, Canada\label{dunlap1}
\and
Dunlap Institute for Astronomy \& Astrophysics, University of Toronto, 50 St. George Street, Toronto, ON M5S 3H4, Canada\label{dunlap2}
\and
Waterloo Centre for Astrophysics, University of Waterloo, Waterloo, ON N2L 3G1, Canada\label{waterloo}
\and
Department of Physics and Astronomy, University of Waterloo, Waterloo, ON N2L 3G1, Canada\label{waterloo2}
\and
Indian Institute of Astrophysics, Koramangala II Block, Bangalore, 560034, India\label{iiabangalore}
%\and
%Department of Astrophysical Sciences, Princeton University, 4 Ivy Lane, Princeton, NJ 08540\label{princeton}
%\and
%Laboratoire Astroparticule et Cosmologie (APC), Université Paris-Cité, Paris, France\label{apc}
%\and
%Department of Physics, University of California, Berkeley, Berkeley, CA 94720, USA\label{berkeley}
}
%\authorrunning{From BeyondPlanck to Cosmoglobe}
\authorrunning{Eskilt et al.}
\titlerunning{CMB parameter estimation}

\abstract{
  We implement support for a cosmological parameter estimation algorithm in \commander\ and quantify its computational efficiency and cost. For a semi-realistic simulation similar to \Planck\ LFI 70\,GHz, we find that the computational cost of producing one single sample is about 20\,CPU-hours and that the typical Markov chain correlation length is $\sim$\,100 samples. The net effective cost per independent sample is $\sim$\,2\,000\,CPU-hours, in comparison with all low-level processing costs of 812\,CPU-hours for \Planck\ LFI and WMAP in \cosmoglobe\ Data Release 1. Thus, although technically possible to run already in its current state, future work should aim to reduce the effective cost per independent sample by one order of magnitude to avoid excessive runtimes, for instance through multi-grid preconditioners and/or derivative-based Markov chain sampling schemes. This work demonstrates the computational feasibility of true Bayesian cosmological parameter estimation with end-to-end error propagation for high-precision CMB experiments without likelihood approximations, but it also highlights the need for additional optimizations before it is ready for full production-level analysis. 
}


\maketitle

%\hypersetup{linkcolor=black}
%\tableofcontents
%\hypersetup{linkcolor=red} 

\section{Introduction}
\label{sec:introduction}

High-precision cosmic microwave background (CMB) measurements currently provide the strongest constraints on a wide range of key cosmological parameters \citep[e.g.,][]{planck2016-l06}. Traditionally, these constraints are derived by first compressing the information contained in the high-dimensional raw time-ordered data into pixelized sky maps and angular power spectra \citep[e.g.,][]{bennett2012,planck2016-l02,planck2016-l03}. This compressed dataset is typically summarized in terms of a low-dimensional power spectrum-based likelihood from which cosmological parameters are derived through Markov chain Monte Carlo (MCMC) sampling \citep{cosmomc,planck2016-l05}.

A key element in this pipeline procedure is the so-called likelihood function, $\mathcal{L}(C_{\ell})$, where $C_{\ell}$ denotes the angular CMB power spectrum \citep[e.g.,][]{planck2016-l05}. This function summarizes both the best-fit values of all power spectrum coefficients and their covariances and correlations. The accuracy of the resulting cosmological parameters, both in terms of best-fit values and their uncertainties, corresponds directly to the accuracy of this likelihood function. As such, this function must account for both low-level instrumental effects (such as correlated noise and calibration uncertainties) and high-level data selection and component separation effects. In general, this function is neither factorizable nor Gaussian, and, except for a few well known special cases, it has no analytical exact and closed form. Unsurprisingly, significant efforts have therefore been devoted to establishing computationally efficient and accurate approximations. Perhaps the most well known example of this is a standard multivariate Gaussian distribution in units of the angular power spectrum of the sky signal, $C_{\ell}$, with a covariance matrix tuned by end-to-end simulations \citep[e.g.,][]{planck2016-l05}. A second widely used case is that of a low-resolution multivariate Gaussian defined in pixel space with a dense pixel-pixel noise covariance matrix \citep{hinshaw2012,planck2016-l05}. Yet other examples include log-normal or Gaussian-plus-lognormal distributions \citep[e.g.,][]{verde2003}, as well as various hybrid combinations \citep[e.g.,][]{gjerlow2013}. 

However, as the signal-to-noise ratios of modern CMB data sets continue to increase, the relative importance of systematic uncertainties grows, and they are often highly nontrivial to describe through low-dimensional and simplified likelihood approximations, both in terms of computational expense and accuracy. These difficulties were anticipated more than two decades ago, and an alternative end-to-end Bayesian approach was proposed independently by \citet{jewell2004} and \citet{wandelt2004}. The key aspect of this framework is that the analysis is global, meaning all aspects of the data are modeled and fitted simultaneously, as opposed to compartmentalized, which is the more traditional CMB pipeline approach. Technically speaking, this is implemented in the form of an MCMC sampler in a process that is statistically analogous to the very last cosmological parameter estimation step in a traditional pipeline, such as \cosmomc\ \citep{cosmomc} or \cobaya\ \citep{Torrado:2020dgo}, with the one fundamental difference that all parameters in the entire analysis pipeline are sampled over jointly within the MCMC sampler. Thus, while a standard cosmological parameter code typically handles a few dozen, or perhaps a hundred, free parameters, the global approach handles billions of free parameters. One practical method of dealing with such a large number of parameters of different types is Gibbs sampling, which draws samples from a joint distribution by iterating over all relevant conditional distributions.

In principle, a global approach is preferable, since better results are in general obtained by fitting correlated parameters jointly rather than separately. However, and equally clearly, this approach is also organizationally and technically more complicated to implement in practice, simply because all aspects of the analysis have to be accounted for simultaneously within one framework. One example of this type of approach is \commanderthree\ \citep{bp03}, which is an end-to-end CMB Gibbs sampler developed by the \citet{bp01} to reanalyze the \Planck\ Low Frequency Instrument (LFI) observations. To date, using Gibbs sampling within this framework is the only way to sample over the entire joint distribution of low-level instrumental parameters, component maps, and cosmological parameters. This work was subsequently superseded by the \cosmoglobe\ project,\footnote{\url{https://cosmoglobe.uio.no}} which is a community-wide open science collaboration that ultimately aims to perform the same type of analysis for all available state-of-the-art large-scale radio, microwave, and submillimeter experiments to then use the resulting analysis to derive one global model of the astrophysical sky. The first \cosmoglobe\ Data Release (DR1) was made public in March 2023 \citep{watts2023_dr1}, and included the first joint end-to-end analysis of both \Planck\ LFI and the Wilkinson Microwave Anisotropy Probe (WMAP; \citealp{bennett2012}), resulting in lower systematic residuals in both experiments.

There are many other methods of Bayesian inference that can be used to estimate cosmological parameters from CMB maps. For example, \citet{Anderes:2014foa} show how to efficiently sample the joint-posterior of the CMB lensing potential and sky map using Hamiltonian Monte Carlo techniques, and \citet{Carron:2018lcr} created a framework to measure the tensor-to-scalar ratio using an approximate, analytic marginalization of lensed CMB maps. The marginal unbiased score expansion method has also been shown to be a powerful algorithm for high-dimensional CMB analyses \citep{Millea:2021had}.

The numerical challenges involved in implementing cosmological parameter estimation without likelihood approximations in Gibbs sampling were originally explored and partially resolved by \citet{jewell:2009} and \citet{racine:2016}. A similar approach was taken by \citet{Millea:2020cpw}, who derived a re-parameterization of the unlensed sky map and the lensing potential to improve the runtime of Gibbs sampling, similar to the method of \citet{racine:2016}. This framework was later enhanced in \citet{Millea:2020iuw} to included several systematic parameters and was applied to SPTpol data.

In principle, each of the aforementioned methods could be used as the CMB map-to-parameter sampling step within a larger time-ordered-data-to-parameter end-to-end Gibbs sampler. The goal of this work is to focus on the algorithm of \citet{racine:2016} and implement it in \commanderthree. We will apply it to simulations with realistic noise and sky coverages, validate the implementation and evaluate its performance. Until now, the highest-level output from \commanderthree\ has been the angular CMB power spectrum. With the code development that led to this paper, it is now technically able to also directly output cosmological parameters.

The rest of the paper is organized as follows. In Sect.~\ref{sec:methods} we provide a brief review of both the \cosmoglobe\ Gibbs sampler method in general and the specific cosmological parameter sampler described by \citet{racine:2016}. In Sect.~\ref{sec:results} we validate our implementation by comparing it to two special cases, namely (i) \cobaya\ coupled to a uniform and (ii) full-sky case and a special-purpose Python sampler for a case with uniform noise but a constant latitude Galactic mask. We also characterize the computational costs of the new sampling step for various data configurations. Finally, we summarize and conclude in Sect.~\ref{sec:conclusions}.

\section{Algorithms}
\label{sec:methods}

We start by briefly reviewing the Bayesian framework for global end-to-end analysis as described by the \citet{bp01} and \citet{watts2023_dr1} and the cosmological parameter estimation algorithm proposed by \citet{racine:2016}, and discuss how these can be combined in \commanderthree\ \citep{bp03}.

\subsection{Bayesian end-to-end CMB analysis}

The main algorithmic goal of the \cosmoglobe\ framework is to derive a numerical representation of the full posterior distribution $P(\omega\mid\d)$, where $\omega$ is the set of all free parameters and $\d$ represents all available data. In this notation, $\omega$ simultaneously accounts for instrumental, astrophysical, and cosmological parameters, which are typically explicitly defined by writing down a signal model, for instance taking the form
\begin{equation}
    \label{eq:data_model}
    \d = \s(\omega) + \n,
\end{equation}
where $\d$ is the data vector; $\s(\omega)$ represents some general data model with free parameters $\omega$; and $\n$ is instrumental noise. In practice, one typically assumes that the noise is Gaussian distributed with a covariance matrix $\N$, and the likelihood can therefore be written as
\begin{equation}
  \mathcal{L}(\omega) = P(\d\mid\omega) \propto e^{-\frac{1}{2}\left(\d-\s(\omega)\right)^T\N^{-1}\left(\d-\s(\omega)\right)}.
\end{equation}
The full posterior distribution reads $P(\omega\mid\d) \propto \mathcal{L}(\omega)P(\omega)$, where $P(\omega)$ represents some set of user-defined priors.

As a concrete example, \cosmoglobe\ Data Release 1 \citep{watts2023_dr1} adopted the following data model for the time-ordered data to describe \Planck\ LFI and WMAP,
\begin{equation}
	\label{eq:model}
	\d =\tens{G}\P\B\M\a+ \s^\mathrm{orb}
	+\s^\mathrm{fsl} + \s^\mathrm{inst}+ \n^\mathrm{corr}+\n^\mathrm{w},
\end{equation}
where $\mathsf G$ is a time-dependent gain factor; $\mathsf P$ is a pointing matrix;
$\B$ denotes instrumental beam convolution; $\mathsf M$ is a mixing matrix that describes the amplitude of a given sky component at a given frequency; $\a$ describes the amplitude of each component at each point in the sky; $\s^\mathrm{orb}$ is the orbital CMB dipole; $\s^\mathrm{fsl}$ is a far sidelobe contribution; $\s^\mathrm{inst}$ is an instrument-specific correction term; $\n^\mathrm{corr}$ denotes correlated noise; and $\n^\mathrm{w}$ denotes white noise. This expression directly connects important instrumental effects (e.g., gains, beams, and correlated noise) with the astrophysical sky (e.g., $\M$ and $\a$), and provides a well-defined model for the data at the lowest level; this approach is what enables global modeling.

For our purposes, the most important sky signal component is the CMB anisotropy field, $\a_{\mathrm{CMB}}$. The covariance of this component reads $\S = \langle \a_{\mathrm{CMB}}^T \a_{\mathrm{CMB}}\rangle$. Under the assumption of a statistically isotropic universe, this matrix is diagonal in harmonic space, $\langle a_{\ell m} a^*_{\ell' m'}\rangle = C_{\ell}(\theta) \delta_{\ell\ell'}\delta_{mm'}$, where $a_{\mathrm{CMB}}(\hat{n}) = \sum_{\ell,m} a_{\ell m} Y_{\ell m}(\hat{n})$ is the usual spherical harmonics expansion and $C_{\ell}$ is called the angular power spectrum. This power spectrum depends directly on a small set of cosmological parameters, $\theta$, and for most universe models can be computed efficiently by Boltzmann solvers such as \camb\ \citep{Lewis:1999bs} or \texttt{CLASS} \citep{lesgourgues:2011}.

With this notation, the goal is now to compute $P(\omega\mid\d)$, where $\omega = \{\tens{G}, \n_{\mathrm{corr}}, \M, \a, \theta, \ldots\}$. Unfortunately, directly evaluating or sampling from this distribution is unfeasible. In practice, we therefore resort to MCMC sampling in general, and Gibbs sampling in particular, which is a well-established method for sampling from a complicated multivariate distribution by iterating over all conditional distributions. For the data model described in Eq.~\eqref{eq:model}, this process can be described schematically in the form of a Gibbs chain, 
  \begin{alignat}{9}
    \label{eq:gain_samp_dist}\tens{G} &\,\leftarrow          P(\tens{G}&\,               \mid \data, &\,\phantom{\tens{G},} &\,\ncorr,&\,\M, &\,\a, &\,\theta)\\
    \label{eq:ncorr_samp_dist} \ncorr &\,\leftarrow    P(\ncorr&\,        \mid \data, &\,\tens{G}, &\,\phantom{\ncorr,}  &\,\M, &\,\a, &\,\theta)\\
    \label{eq:beta_samp}\M &\,\leftarrow                     P(\M &\, \mid \data, &\,\tens{G}, &\,\ncorr, &\,\phantom{\M}, &\,\a, &\,\theta)\\
    \a &\,\leftarrow                                   P(\a&\,            \mid \data, &\,\tens{G}, &\,\ncorr, &\,\M, &\,\phantom{\a,} &\,\theta)\\
    \theta &\,\leftarrow                             P(\theta &\,         \mid \data, &\,\tens{G}, &\,\ncorr, &\,\M, &\,\a,&\,\phantom{\theta})\label{eq:param_samp},
    \end{alignat}
  where $\leftarrow$ indicates replacing the parameter on the left-hand side with a random sample from the distribution on the right-hand side.

As described by the \citet{bp01} and \citet{watts2023_dr1}, this process has been implemented in \commanderthree\ by \BP\ and \Cosmoglobe. However, in the code described there, the last step only supports a power spectrum estimation (i.e., $\theta = C_{\ell}$). In this work, we did not sample any instrumental or foreground parameters, and we assumed a foreground-cleaned map of the CMB as a starting point. Rather, the goal of the paper is to replace the last step, Eq.~\eqref{eq:param_samp}, with actual cosmological parameter estimation of a CMB map, in which $\theta$ takes on the usual form of the dark matter density $\Omega_\mathrm{c}$, the Hubble constant $H_0$, the spectral index of scalar perturbations $n_\mathrm{s}$, etc.

\subsection{Joint sampling of CMB sky signal and cosmological parameters}

Cosmological parameter estimation through end-to-end CMB Gibbs sampling has been discussed in the literature for almost two decades, starting with \citet{jewell2004} and \citet{wandelt2004}. However, as pointed out by \citet{eriksen:2004}, a major difficulty regarding this method is a very long correlation length in the low signal-to-noise regime, i.e., at high multipoles. Intuitively, the origin of this problem lies in the fundamental Gibbs sampling algorithm itself, namely that it only allows for parameter variations parallel or orthogonal to the coordinate axes of each parameter in question, and not diagonal moves. For highly degenerate distributions, this makes it very expensive to move from one tail of the distribution to the other. A solution to this problem was proposed by \citet{jewell:2009}, who introduced a joint CMB sky signal and $C_\ell$ move. This idea was further refined by \citet{racine:2016}, who noted that faster convergence could be obtained by only rescaling the low signal-to-noise multipoles. In the following, we give a brief summary of these ideas, both standard Gibbs sampling in Sect.~\ref{sec:gibbs} and joint sampling in Sect.~\ref{sec:joint-sampling}.

\subsubsection{Standard CMB Gibbs sampling}
\label{sec:gibbs}

Starting with the standard Gibbs sampling algorithm \citep{jewell2004,wandelt2004}, we first note that as far as cosmological parameter estimation with CMB data is concerned, only the CMB map, $\a_{\mathrm{CMB}}$ (which we for brevity denote $\a$ in the rest of this section), is directly relevant. All other parameters in Eq.~\eqref{eq:model} only serve to produce the cleanest possible estimate of $\a$ and to propagate the corresponding uncertainties. From now on, we therefore consider a greatly simplified data model of the form,
\begin{equation}
  \r = \B\a + \n,
\end{equation}
where $\r$ now represents a foreground-cleaned CMB map (or residual) obtained by subtracting all other nuisance terms from the raw data, and our goal is to estimate cosmological parameters from this map. Under the assumption that the CMB is a Gaussian random field, this could in principle be done simply by mapping out the exact marginal posterior distribution by brute force,
\begin{equation}
  P(\theta \mid \r) \propto \mathcal{L}(\theta)P(\theta) \propto \frac{e^{-\frac12 \r^T (\B^T \S(\theta) \B + \N)^{-1}\r}}{\sqrt{\left|\B^T \S(\theta) \B + \N\right|}},
  \label{eq:exact_posterior}
\end{equation}
where $\B^T \S(\theta) \B$ is the beam-convolved CMB signal covariance matrix and $\N$ is the corresponding noise covariance matrix, and we have assumed uniform priors, $P(\theta) = 1$. This, however, becomes intractable for modern CMB experiments as the computational expense scales as $\mathcal{O}(N_p^3)$, where $N_p$ is the number of pixels, and nowadays typical CMB maps include millions of pixels.

\begin{figure}
	\centering
	\includegraphics[width=\linewidth]{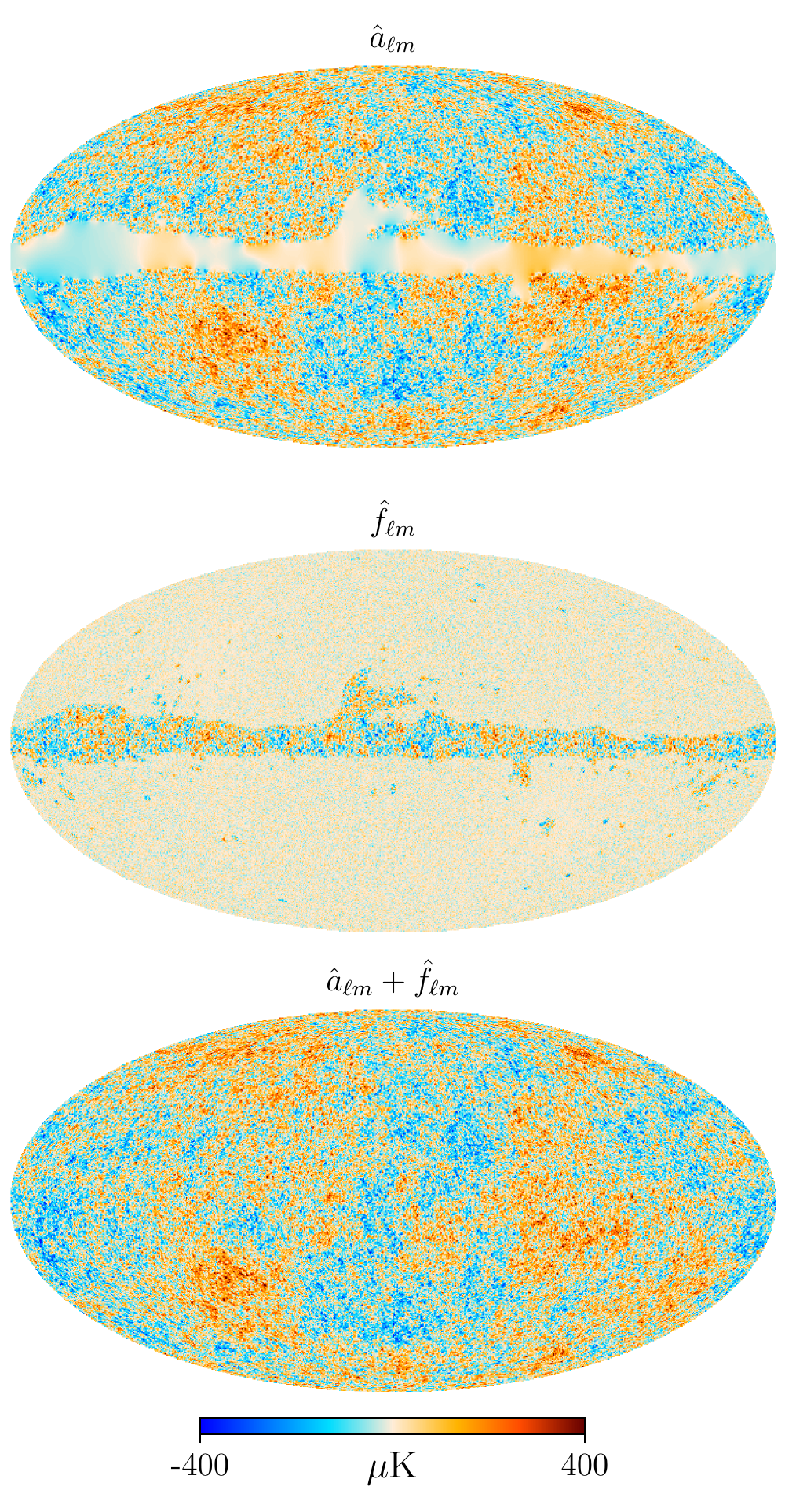}
	\caption{\label{fig:sky_map}Example of a constrained CMB sky map realization produced during traditional Gibbs sampling. From top to bottom, the three panels show the mean field map ($\hat{\a}$), the fluctuation map ($\hat{\f}$), and the full constrained realization ($\a = \hat{\a}+\hat{\f}$). This example is generated with the data configuration discussed in Sect.~\ref{sec:results}, which corresponds to \Planck\ LFI 70\,GHz noise properties and a realistic Galactic mask.}
\end{figure}
\begin{figure}
	\centering
	\includegraphics[width=\linewidth]{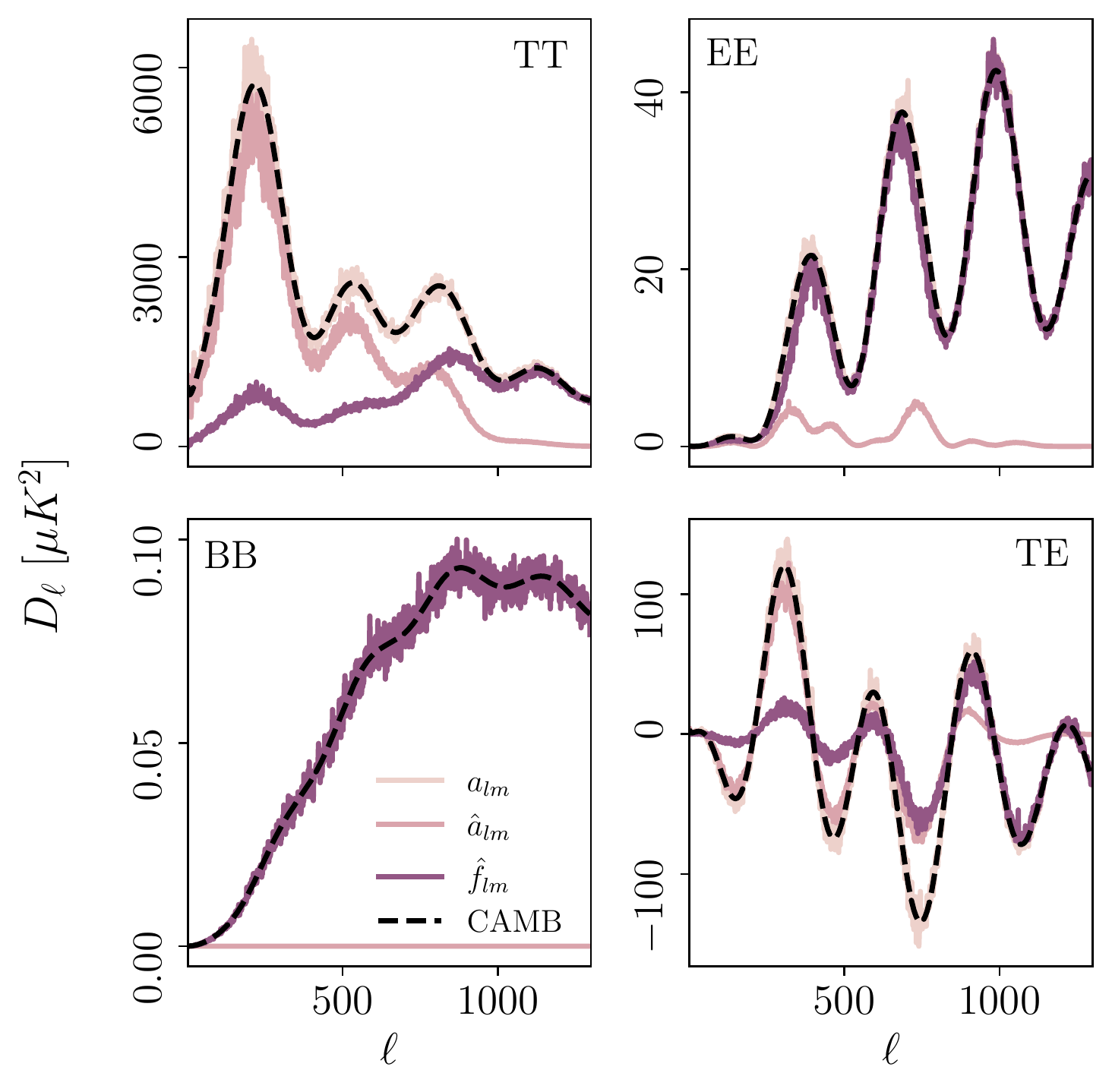}
	\caption{\label{fig:sigma_ell}Angular power spectra for each of the constrained realization maps shown in Fig.~\ref{fig:sky_map}. The $\Lambda$CDM cosmological model corresponding to $\theta$ is shown as a dashed black line, while the colored curves show (from dark to light) spectra for the fluctuation map, the mean field map, and the full constrained realization. Note that in the fully noise-dominated cases, such as the entire B-mode power spectrum and all high multipoles, the fluctuation term $\hat f_{\ell m}$ and the CMB estimate $a_{\ell m}$ have nearly identical power spectra by design.}
\end{figure}

In the special (and unrealistic) case of uniform noise and no Galactic mask, this likelihood can be simplified in harmonic space, in which it does become computationally tractable. Let us define the observed angular CMB power spectrum (including beam convolution and noise) to be $\hat{C}^{\mathrm{o}}_{\ell} = \frac{1}{2\ell+1}\sum_m r_{\ell m}r^*_{\ell m}$ and the corresponding white noise spectrum to be defined by $\N_{\ell m \ell'm'} = N_\ell \delta_{\ell \ell'}\delta_{mm'}$. In that case, the likelihood in Eq.~\eqref{eq:exact_posterior} for temperature-only observations simplifies to
{\small \begin{equation}
  \ln P(\theta \mid \r) = \sum_{\ell} -\frac{2\ell+1}{2} \left[\frac{\hat{C}^{\mathrm{o}}_{\ell}}{b_\ell^2 C_{\ell}(\theta) + N_\ell}-\ln \left(\frac{\hat{C}^{\mathrm{o}}_{\ell}}{b_\ell^2 C_{\ell}(\theta) + N_\ell} \right) \right],
  \label{eq:exact_harm}
\end{equation}}
where $b_\ell$ is the beam response function and we have removed constant terms; a similar expression for polarization is available in, for example, \citet{larson:2006} and \citet{Hamimeche:2008ai}. This expression is used later to validate the production code.

While it is difficult to evaluate Eq.~\eqref{eq:exact_posterior} directly, it is in fact possible to sample from it using Gibbs sampling \citep{jewell2004,wandelt2004}. First, we note that the joint posterior distribution $P(\theta, \a \mid \r)$ can be written as
\begin{align}
    \nonumber
    P(\theta, \a \mid \r) &= \frac{P(\theta, \a, \r)}{P(\r)} = P(\r \mid \a)P(\a\mid \theta)\frac{P(\theta)}{P(\r)}\\
    \label{eq:joint-posterior}
    &\propto \frac{e^{-\frac12 \left(\r-\B\a \right)^T \N^{-1}\left(\r-\B\a \right)}}{\sqrt{\left|\N\right|}}
    \frac{e^{-\frac12 \a^T \S^{-1}\a}}{\sqrt{\left|\S\right|}}\frac{P(\theta)}{P(\r)},
\end{align}
and drawing samples from this distribution can be achieved through Gibbs sampling,
\begin{align}
    \label{eq:a-gibbs}
    \a^{i+1} &\leftarrow P(\a \mid \theta^{i}, \r),\\
    \label{eq:theta-gibbs}
    \theta^{i+1} &\leftarrow P(\theta \mid \a^{i+1}, \r).
\end{align}
With these joint samples in hand, a numerical representation of $P(\theta \mid \r)$ can be obtained simply by marginalizing over $\a$, which is equivalent to simply making histograms of the $\theta$ samples.

For this algorithm to work, we need to be able to sample from each of the two conditional distributions in Eqs.~\eqref{eq:a-gibbs} and \eqref{eq:theta-gibbs}. Starting with the cosmological parameter distribution in Eq.~\eqref{eq:theta-gibbs}, we note that $P(\theta \mid \a^{i+1}, \r) = P(\theta \mid \a^{i+1})$, which is equivalent to the distribution in Eq.~\eqref{eq:exact_harm} with no beam and noise. This can be coupled to a standard Boltzmann solver and Metropolis-Hastings sampler, similar to \camb\ and \cosmomc. We return to this step in the next section.

Next, for the amplitude distribution in Eq.~\eqref{eq:a-gibbs}, one can show by ``completing the square'' of $\a$ in Eq.~\eqref{eq:joint-posterior} that 
\begin{equation}
    P(\a \mid \theta, \r) \propto e^{-\frac12 \left(\a - \hat{\a}\right)^T \left(\S^{-1} + \B^T\N^{-1}\B\right) \left(\a - {\hat{\a}}\right)},
\end{equation}
where we have defined the so-called mean field map,
\begin{equation}
\label{eq:mean-field-map}
\hat{\a} \equiv \left[\S^{-1} + \B^T \N^{-1}\B \right]^{-1} \B^T \N^{-1} \r.
\end{equation}
This is a multivariate Gaussian distribution with mean $\hat{\a}$ and a covariance matrix $\left[\S^{-1} + \B^T \N^{-1}\B \right]^{-1}$ that can be sampled from solving the following equation using conjugate gradients \citep{shewchuk:1994,seljebotn:2019},
\begin{equation}
    \label{eq:mapmakingeq}
    \left[\S^{-1} + \B^T \N^{-1}\B \right]\a = \B^T \N^{-1} \r + \S^{-\frac{1}{2}}\w_0 +\B^T \N^{-\frac{1}{2}}\w_1,
\end{equation}
where $\w_0$ and $\w_1$ are randomly drawn Gaussian maps with unit variance and zero mean. The resulting sample is typically referred to as a ``constrained realization.''

For the purposes of the next section, it is useful to decompose the full constrained realization into two separate components, namely the mean field map defined by Eq.~\eqref{eq:mean-field-map} and the fluctuation map, 
\begin{equation}
\label{eq:fluc-map}
\hat{\f} \equiv \left[\S^{-1} + \B^T \N^{-1}\B \right]^{-1} \left(\S^{-\frac{1}{2}}\w_0 +\B^T \N^{-\frac{1}{2}}\w_1 \right)
\end{equation}
such that $\a = \hat{\a} + \hat{\f}$. For a data configuration with uniform noise and full-sky temperature data, these equations can be solved exactly in spherical harmonics space, fully analogously to Eq.~\eqref{eq:exact_harm},
\begin{align}
    \label{eq:hat_s_approx}
    \hat{a}_{\ell m} &= r_{\ell m}\frac{b_{\ell}C_{\ell}}{N_\ell + b_{\ell}^2C_{\ell}},\\
    \label{eq:hat_f_approx}
    \hat{f}_{\ell m} &= w_{0\ell m}\frac{N_{\ell}\sqrt{C_{\ell}}}{N_\ell + b_{\ell}^2C_{\ell}}+w_{1\ell m}\frac{\sqrt{N_{\ell}}b_{\ell}C_\ell}{N_\ell + b_{\ell}^2C_{\ell}}.
\end{align}
As summarized in 
Appendix~\ref{sec:appendixA}, we can find a closed-form expression for $N_{\ell m\ell'm'}^{-1}$ for a constant latitude mask with isotropic noise which greatly simplifies solving Eq.~\eqref{eq:mapmakingeq}, and we have made a Python script that performs these calculations in order to validate the main \commanderthree\ code discussed below.

To build intuition, these maps are illustrated for one random sample with semi-realistic instrument characteristics in Fig.~\ref{fig:sky_map}. The top panel shows the mean field map (or ``Wiener filter map''); this map summarizes all significant information in $\r$, and each mode is weighted according to its own specific signal-to-noise ratio taking into account both instrumental noise and masking. As a result, all small scales in the Galactic plane are suppressed, and only the largest scales are retained, which are constrained by the high-latitude observations in conjunction with the assumptions of both Gaussianity and statistical isotropy. The second panel shows the fluctuation map, which essentially serves to replace the lost power in $\hat{\a}$ with a completely random fluctuation. The sum of the two terms, shown in the bottom panel, represents one possible full-sky realization that is consistent with both the data and the assumed cosmological parameters, $\theta$. If we were to produce a second such constrained realization, it would look almost identical at high Galactic latitudes, where the data are highly constraining, while the Galactic plane would exhibit different structures on small and intermediate scales.

A corresponding comparison of angular power spectra is shown in Fig.~\ref{fig:sigma_ell}. The relative signal-to-noise ratio of each multipole moment can be seen as the ratio between the intermediate and dark-colored lines. In temperature, this value is higher than unity up to $\ell\lesssim700$, while for $EE$ polarization it is less than unity everywhere except at the very lowest multipoles. A closely related way to interpret this figure is that the intermediate curve shows the total information content of the data. Thus, for the $BB$ spectrum there is no visually relevant information present at all, and the full constrained realization is entirely defined by the fluctuation map, which is given by the assumed cosmological model, $\theta$. This explains why the standard Gibbs sampler works well in the high signal-to-noise regime but poorly in the low signal-to-noise limit: When we have high instrumental noise, $\N \rightarrow \infty$, then we have no information about $a_{\ell m}$ from the data $r_{\ell m}$, and the total sky sample only becomes a realization of the assumed variance, $C_{\ell}$ (i.e., $a_{\ell m} \rightarrow \hat{f}_{\ell m} = \sqrt{C_{\ell}} w_{0\ell m}$). However, since we have no information of the mean field map $\hat{a}_{\ell m} \rightarrow 0$, we also do not know $C_\ell$. Hence, we get a strong degeneracy between the $C_\ell$ and $a_{\ell m}$. Gibbs sampling is well known to perform poorly for strongly correlated parameters, as one cannot move diagonally through the parameter space from one end of the joint posterior distribution to the other. 

\begin{figure}
	\centering
	\includegraphics[width=\linewidth]{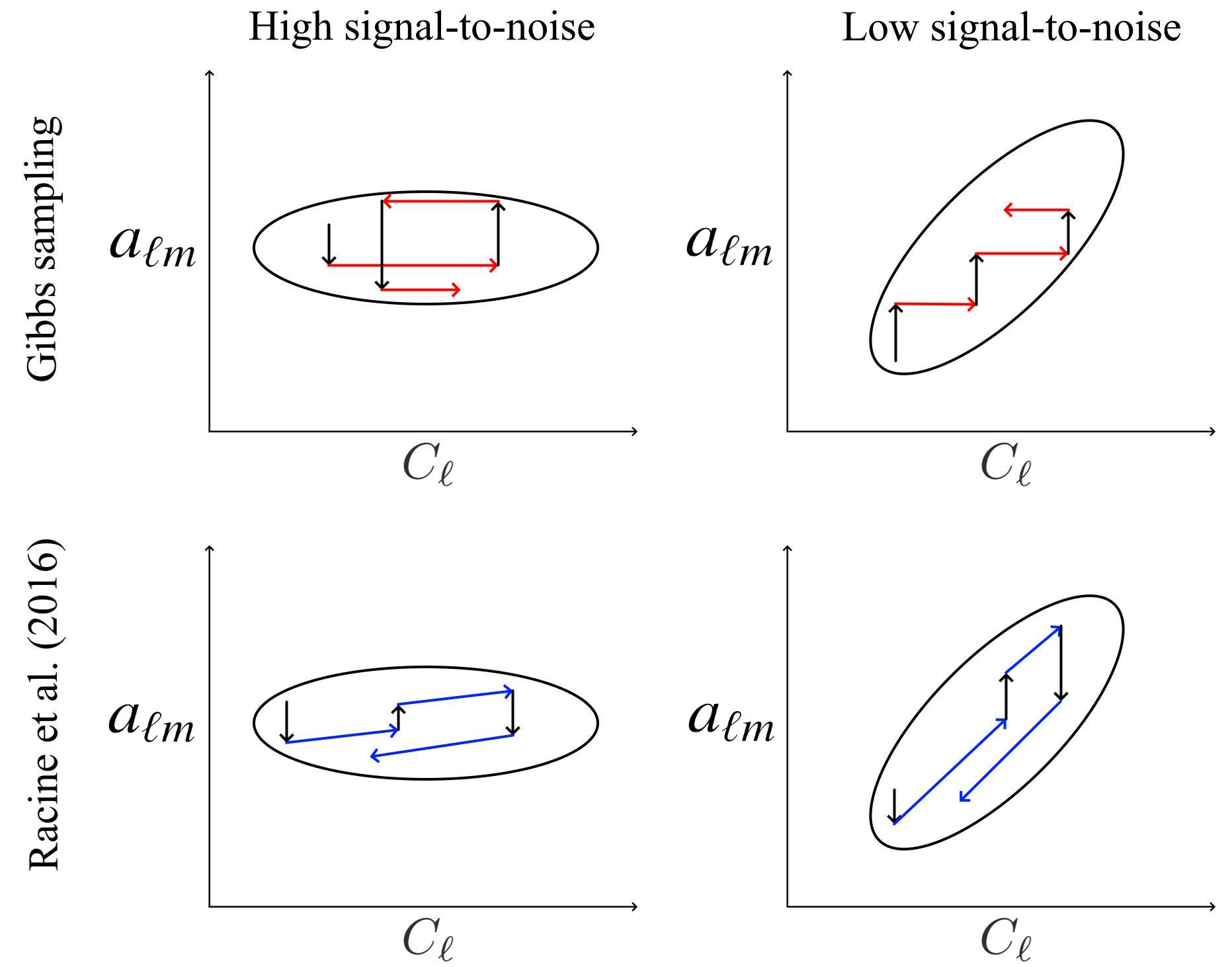}
	\caption{\label{fig:illustration}Schematic comparison of the standard Gibbs sampler (top panels) and the joint sampler of \citet{racine:2016}, bottom panels. The left and right columns show high and low signal-to-noise regimes, respectively. The Gibbs sampler performs poorly in the low signal-to-noise regime as it requires a large number of samples to explore the posterior distribution. The joint sampler in the lower panels performs well in both regimes as it allows the next sample to move diagonally in the $\{\a, \theta\}$ parameter space.}
\end{figure}

\subsubsection{Joint sky signal and cosmological parameter sampling}
\label{sec:joint-sampling}

As the Gibbs sampler struggles with long correlation lengths, a joint sampler was proposed by \citet{jewell:2009} and \citet{racine:2016} that was designed to establish an efficient sampling algorithm for the degeneracy between $\a$ and $\theta$ in the low signal-to-noise regime. Mathematically, this algorithm is a standard Metropolis-Hastings sampler with a Gaussian transition rule for $\theta$,
\begin{equation}
w(\theta \mid\theta^i) = e^{-\frac12 \left(\theta - \theta^i \right)^T \tens{C}_{\theta}^{-1}\left(\theta - \theta^i \right)},
\end{equation}
where $\tens{C}_{\theta}$ is a user-specified (and pre-tuned) proposal covariance matrix made from a previous chain using a suboptimal, diagonal proposal matrix. This is followed by a deterministic rescaling of the fluctuation map in the signal amplitude map,
\begin{equation}
    a_{\ell m}^{i+1} = \hat{a}_{\ell m}^{i+1} + \left(\tens{C}^{i+1}_{\ell}\right)^{1/2}\left(\tens{C}^{i}_{\ell}\right)^{-1/2} \hat{f}_{\ell m}^{i},
\end{equation}
where we have defined the covariance matrix
\begin{equation}
  \tens{C}^i_\ell = \begin{bmatrix}
    C^{TT}_{\ell}\left(\theta^i\right) & C^{TE}_\ell\left(\theta^i\right) & 0\\
    C^{TE}_{\ell}\left(\theta^i\right) & C^{EE}_\ell\left(\theta^i\right) & 0\\
    0 & 0 & C^{BB}_\ell\left(\theta^i\right).
  \end{bmatrix}
\end{equation}
Here, we have set the parity-odd power spectra to zero as predicted by $\Lambda$ cold dark matter (CDM), $C^{EB}_\ell = C^{TB}_\ell = 0$. Therefore, we defined the rescaled fluctuation map,
\begin{equation}
  \hat{f}_{\ell m}^{\textrm{scaled},\, i+1} \equiv \left(\tens{C}^{i+1}_{\ell}\right)^{1/2}\left(\tens{C}^{i}_{\ell}\right)^{-1/2} \hat{f}_{\ell m}^{i}.
\end{equation}

As shown by \citet{racine:2016}, the acceptance rate for this joint move is
\begin{equation}
    \label{eq:acceptance-rate}
    A = \mathrm{min}\left[1, \frac{\pi(\theta^{i+1})}{\pi(\theta^i)} \frac{P(\theta^{i+1})}{P(\theta^i)} \right],
\end{equation}
where $P(\theta)$ is the prior on $\theta$ and
\begin{align}
    \nonumber
    \pi(\theta^{i}) = \mathrm{exp}\bigg[&-\frac12 \left(\r-\B\hat{\a}^i\right)^T \N^{-1}\left(\r-\B\hat{\a}^i\right)\\
    &-\frac12\a^{i,T} \S^{i, -1}\hat{\a}^i -\frac12 \hat{\f}^{i, T}\B^T\N^{-1} \B\hat{\f}^i\bigg],
\end{align}
where we for brevity have omitted the $\theta^i$ dependence of $\hat{\a}^i$, $\hat{\f}^i$ and $\S^i$. 

We note that the acceptance rate uses the scaled fluctuation term, $\hat{f}_{\ell m}^{\textrm{scaled},\, i+1}$, for sample $i+1$ instead of the calculated fluctuation term in Eq.~\eqref{eq:mapmakingeq}. Hence, we only need to calculate $f_{\ell m}^{i+1}$ from this equation if the sample is accepted. This allows us to save roughly half the computational time of discarded samples as compared to accepted samples.

The full algorithm can now be summarized in terms of the following steps:
\begin{enumerate}
    \item We start with an initial guess of the cosmological parameters, $\theta^0$. From that, we calculate $\S^0$ with \camb; the mean field map, $\hat{\a}^0$ from Eq.~\eqref{eq:mean-field-map}; and the fluctuation map, $\hat{\f}^0$, from Eq.~\eqref{eq:fluc-map}.
    \item We then draw a new cosmological parameter sample from $w(\theta\mid\theta^i)$, and reevaluate $\S^{i+1}$, $\hat{\a}^{i+1}$, and the scaled fluctuation term $\hat{\f}^{\textrm{scaled},\, i+1}$.
    \item We calculate the acceptance term of Eq.~\eqref{eq:acceptance-rate}, and accept or reject according to the regular Metropolis-Hasting rule.
    \item If the sample is accepted, then we calculate $\hat{\f}^{i+1}$. In the next iteration this term becomes $\hat{\f}^{i}$, which will appear in the acceptance probability and the equation for $\hat{\f}^{\textrm{scaled},\, i+1}$. 
    \item Iterate 2--4.
\end{enumerate}

The intuition behind this joint move is illustrated in Fig.~\ref{fig:illustration}. The top panel shows the standard Gibbs sampling algorithm, in which only parameter moves parallel to the axes are allowed. This works well in the high signal-to-noise regime (left column), where the joint distribution is close to symmetric. In the low signal-to-noise regime, however, the distribution is strongly tilted, because of the tight coupling between $\hat{\f}$ and $\theta$ discussed at the end of Sect.~\ref{sec:gibbs}, and many orthogonal moves are required in order to move from one tail of the distribution to the other. On the other hand, with the joint algorithm, in the bottom panel, steps are non-orthogonal, with a slope defined exactly by the local signal-to-noise ratio of the mode in question. The net result is a much faster exploration of the full distribution.

\begin{figure}
	\centering
	\includegraphics[width=\linewidth]{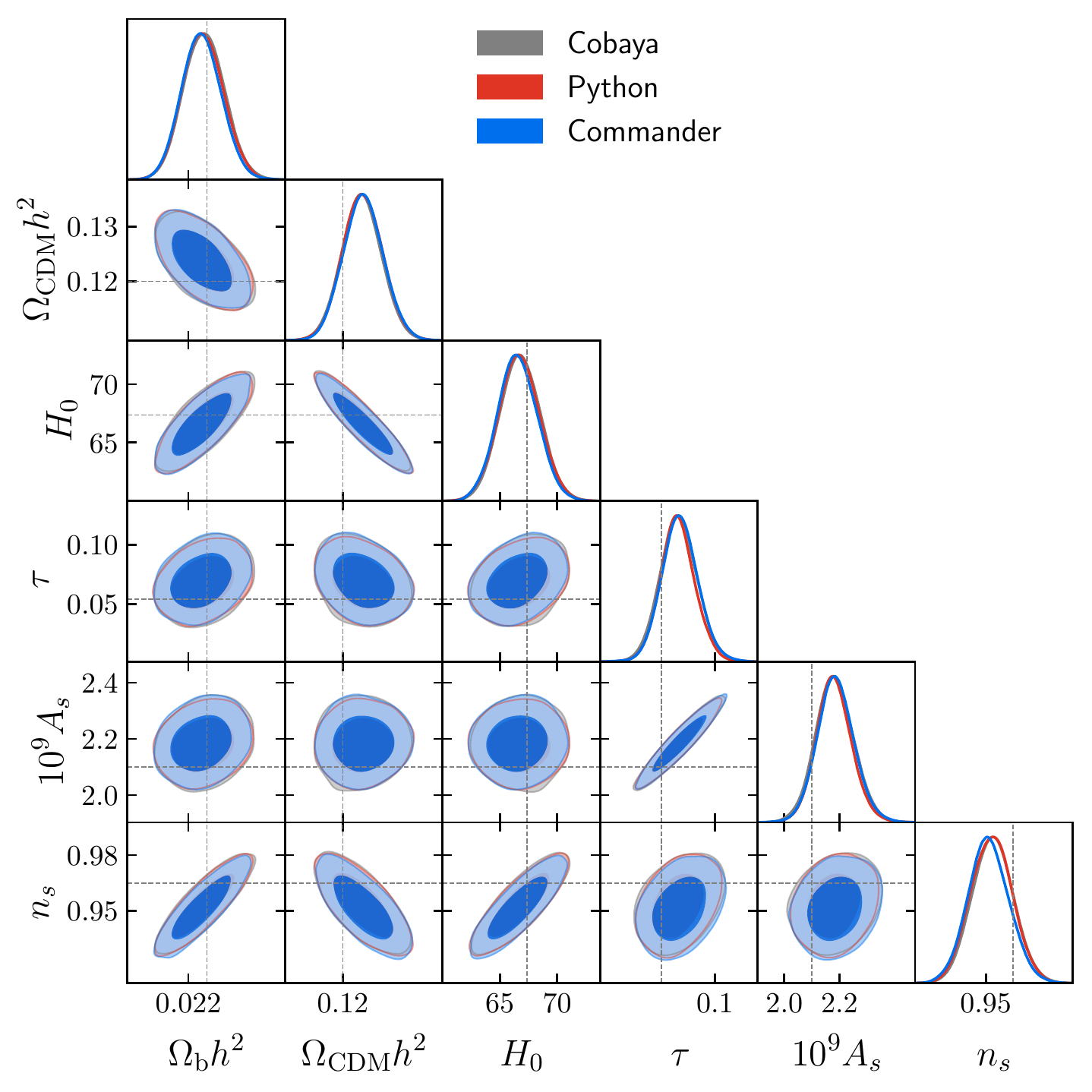}
	\caption{\label{fig:nomask}Comparison of cosmological parameter posterior distributions derived with \commanderthree\ (blue), Python (red), and \cobaya\ (gray) for a simulation with uniform noise and full-sky coverage. They are made from a single chain per code, and the true input cosmological parameter values are indicated by vertical and horizontal dashed lines.}
\end{figure}

\section{Results}
\label{sec:results}

The main goal of this paper is to implement the joint sampling algorithm into \commanderthree\ and characterize its performance both in terms of accuracy and computational cost on simulated data. All \commander\ code used in the following is available as Open Source software in a GitHub repository\footnote{\url{https://github.com/Cosmoglobe/Commander}}. For the current paper, we focused on a standard six-parameter $\Lambda$CDM model and chose $\theta=(\Omega_{\textrm{b}}h^2, \Omega_{\textrm{CDM}}h^2, H_0, \tau, A_s, n_s)$ as our base parameters, where $\Omega_\mathrm{b}$ and $\Omega_\mathrm{CDM}$ are the current density of baryonic and CDM; $H_0$ is the Hubble parameter (and $h$ is the normalized Hubble constant, $h=\frac{H_0}{100\,\mathrm{km s}^{-1} \mathrm{Mpc}^{-1}}$); $\tau$ is the optical depth at reionization; and $A_s$ and $n_s$ are the amplitude and tilt of the scalar primordial power spectrum. We used \camb\footnote{\url{https://github.com/cmbant/CAMB}} \citep{Lewis:1999bs} to evaluate all $\Lambda$CDM power spectra, $\tens{C}_{\ell}(\theta)$.

In the following we consider three different simulations, corresponding to increasing levels of complexity and realism. A natural early target for this algorithm is a reanalysis of \cosmoglobe\ DR1; hence, all three cases are similar to the \Planck\ LFI channels. We use the \planck\ data release 4 70\,GHz beam transfer functions \citep{planck2020-LVII}, and for the two first cases, the noise level is isotropic with a value matching twice the mean of the full-sky 70\,GHz rms map, while for the third case is given by one realization of the actual non-isotropic \BP\ 70\,GHz noise rms distribution.

The second difference between the three cases is their sky coverage. The first case considers full-sky data, for which the analytical posterior distribution solution is readily available through Eq.~\eqref{eq:exact_harm}. The second case implements a constant-latitude Galactic mask for which an analytical expression is also available, albeit more complicated than for the full sky (see Appendix~\ref{sec:appendixA} for details). The third case uses a realistic Galactic and point-source mask, for which no analytical expression is available. 

To validate the main \commanderthree\ results, we developed a special-purpose Python program\footnote{Publicly available at \url{https://github.com/LilleJohs/COLOCOLA}} that implements the same joint sampling algorithm described above for the two first cases, and this serves essentially as a basic bug-check with respect to the Fortran implementation. In addition, we also implemented a uniform, full-sky likelihood module for \cobaya\ \citep{Torrado:2020dgo}, an industry-standard cosmological parameter Metropolis sampler, and this serves as an end-to-end check on the entire code, including integration with \camb.

\subsection{Accuracy validation}

\begin{figure}
	\centering
	\includegraphics[width=\linewidth]{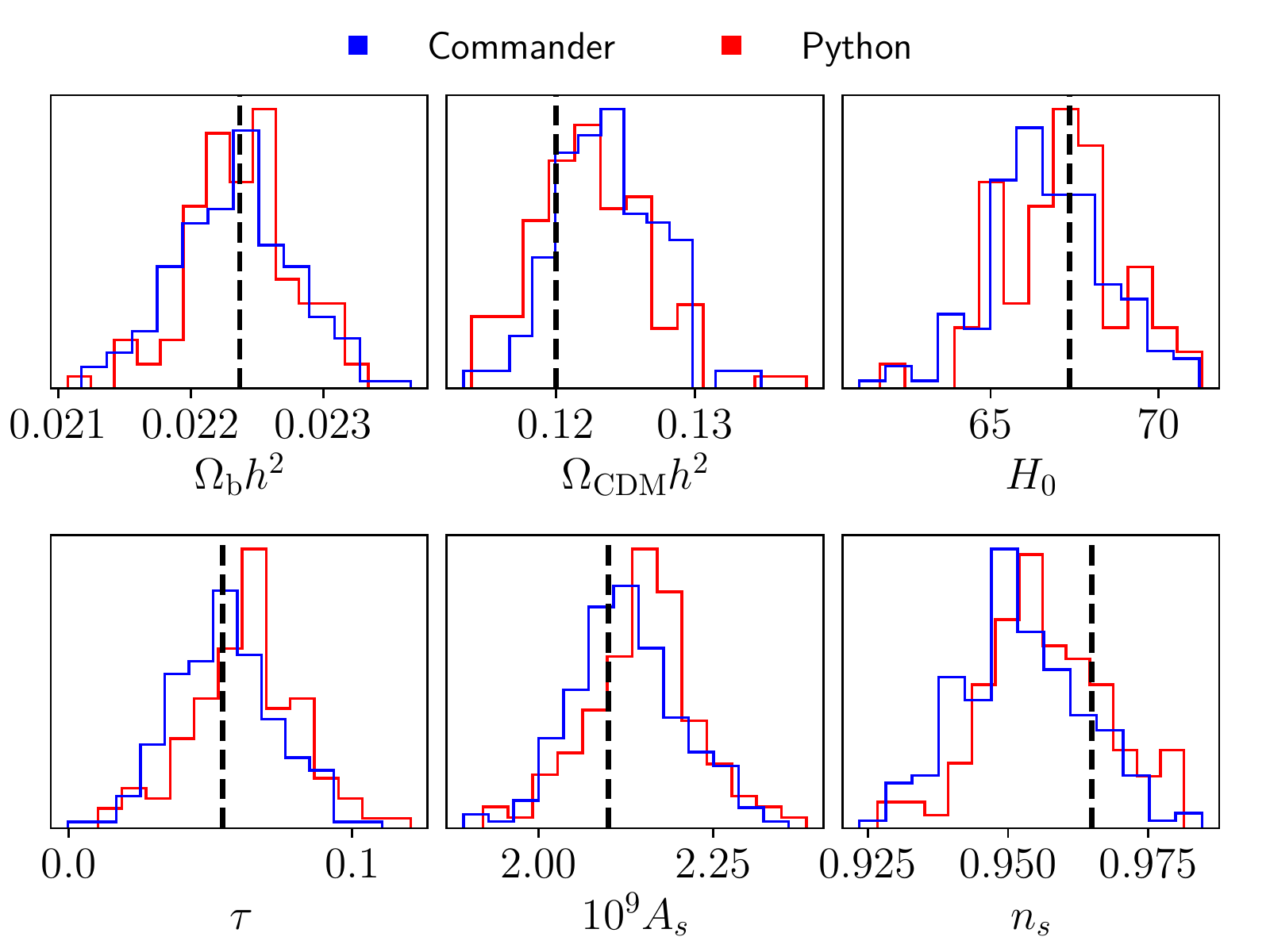}
	\caption{\label{fig:mask10}Comparison of marginal cosmological parameter posterior distributions derived with \commanderthree\ and Python for a simulation with uniform noise and a 10\,\% constant latitude mask. The total number of independent samples produced by the two codes is $\mathcal{O}(10^2)$, which accounts for the Markov chain correlation length of the algorithm; the Monte Carlo uncertainty due to a finite number of samples is therefore significant.}
\end{figure}

\begin{figure}
	\centering
	\includegraphics[width=\linewidth]{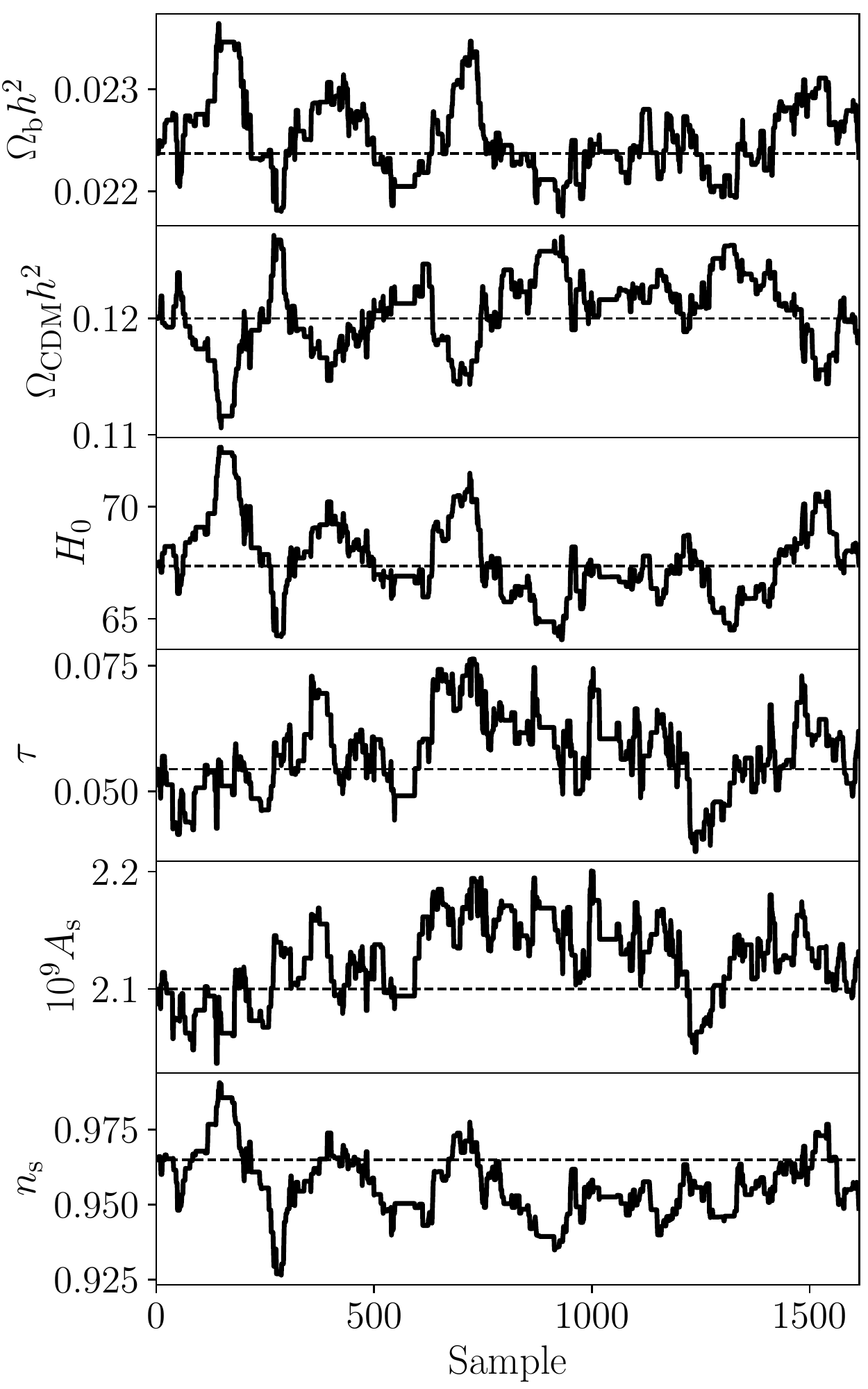}
	\caption{\label{fig:traceplot}Trace plot of the cosmological parameters for a semi-realistic case based on the \Planck\ LFI $70\,$GHz noise level and a realistic Galactic mask. The dashed horizontal lines show the true input values.}
\end{figure}

The first main goal of this paper is to demonstrate that the new \commanderthree\ implementation of the \citet{racine:2016} algorithm performs as expected with respect to accuracy. Thus, we start our discussion by considering the uniform and full-sky case discussed above. For each of the three available codes - \commanderthree, Python, and \cobaya - we produced 50\,000 samples; we show the full posterior distributions in Fig.~\ref{fig:nomask}. The agreement between all three codes is excellent, and the overall variations are within $\sim$\,0.1$\,\sigma$.

Second, Fig.~\ref{fig:mask10} shows a similar comparison between the \commanderthree\ and Python implementation. The sky mask is in this case defined by $|b|<5.7^{\circ}$ where $b$ is the Galactic latitude, removing 10\,\% of the sky. We also find good agreement in this case between the two implementations, and the true input values (marked by vertical dashed lines) all fall well within the full posterior distributions. However, we do note that the introduction of a sky mask significantly increases the overall run time of the algorithm, since one now has to solve for the mean field map repeatedly at each step with conjugate gradients, rather than with brute force matrix inversions. In total, we only produce $\sim$\,11\,000 and 17\,000 samples for each of the two codes for this exercise, at a total cost of $\mathcal{O}(10^5)$ CPU-hours.

Finally, in Fig.~\ref{fig:traceplot} we show trace plots for each cosmological parameter for the realistic configuration as produced with \commanderthree, noting that none of the other codes are technically able to produce similar estimates. In this case, the computational cost is even higher, and we only produce a total of 1\,600 samples. Still, by comparing these traces with the true input values (marked as dashed horizontal lines), we do see that the resulting samples agree well with the expected mean values. The uncertainties are also as expected since they are smaller and on the order of the noisier 10\% constant latitude and isotropic noise case in Fig.~\ref{fig:mask10}.

\subsection{Markov chain correlation length and computational costs}
\label{sec:resources}

\begin{figure}
	\centering
	\includegraphics[width=\linewidth]{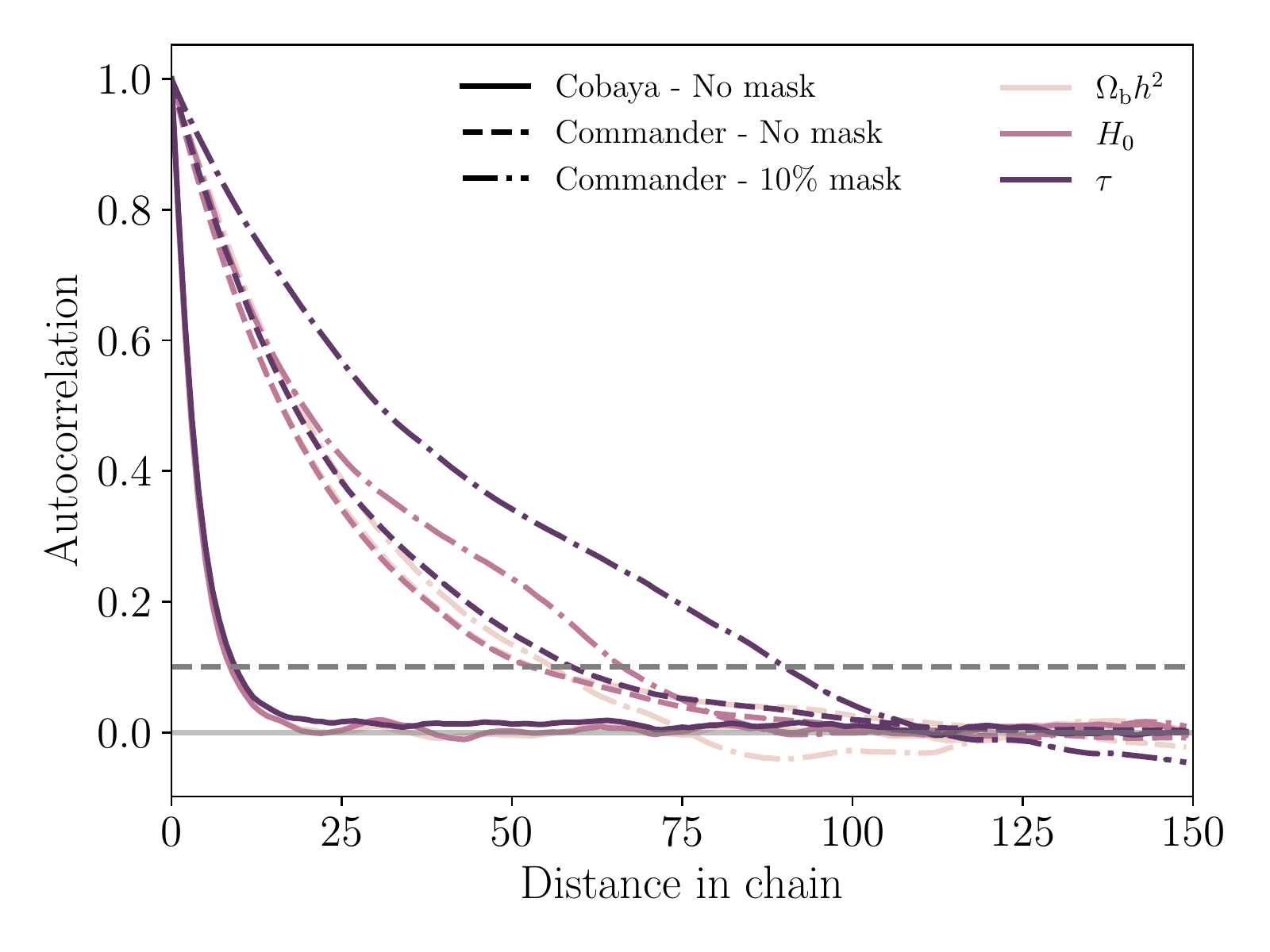}
	\caption{\label{fig:autocorrelation}Markov chain autocorrelation functions for \cobaya\ (solid lines) and \commanderthree\ for a set of typical cosmological parameters. Dashed and dot-dashed lines show \commanderthree\ results for the full-sky case and the 10\,\% mask case, respectively. The horizontal dashed gray line indicates an autocorrelation of 0.1, which we use to define the overall correlation length.   }
\end{figure}

The second main goal of this paper is to quantify the computational costs involved in this algorithm and identify potential bottlenecks that could be optimized in future work. The effective cost per independent Markov chain sample can be written as the product of the raw sampling cost per sample and the overall Markov chain correlation length. The latter can be quantified in terms of the autocorrelation function,
\begin{equation}
  \zeta_\theta(\Delta) \equiv \left<\frac{(\theta_i - \mu)}{\sigma}\frac{(\theta_{i+\Delta} - \mu)}{\sigma}\right>,
\end{equation}
where $\mu$ and $\sigma$ are the posterior mean and standard deviation computed from the Markov chain for $\theta$. We define the overall correlation length to be the first value of $\Delta$ for which $\zeta < 0.1$. This function is plotted for both \cobaya\ and \commanderthree\ in Fig.~\ref{fig:autocorrelation}; for the latter, dashed lines show full-sky results and dot-dashed lines show results for a 10\,\% mask. Overall, we see that the \commanderthree\ correlation length is 50--60 for the full-sky case, increasing to 50--100 for the masked case. For comparison, \cobaya\ typically has a correlation length of about 10.

\begin{figure}
	\centering
	\includegraphics[width=\linewidth]{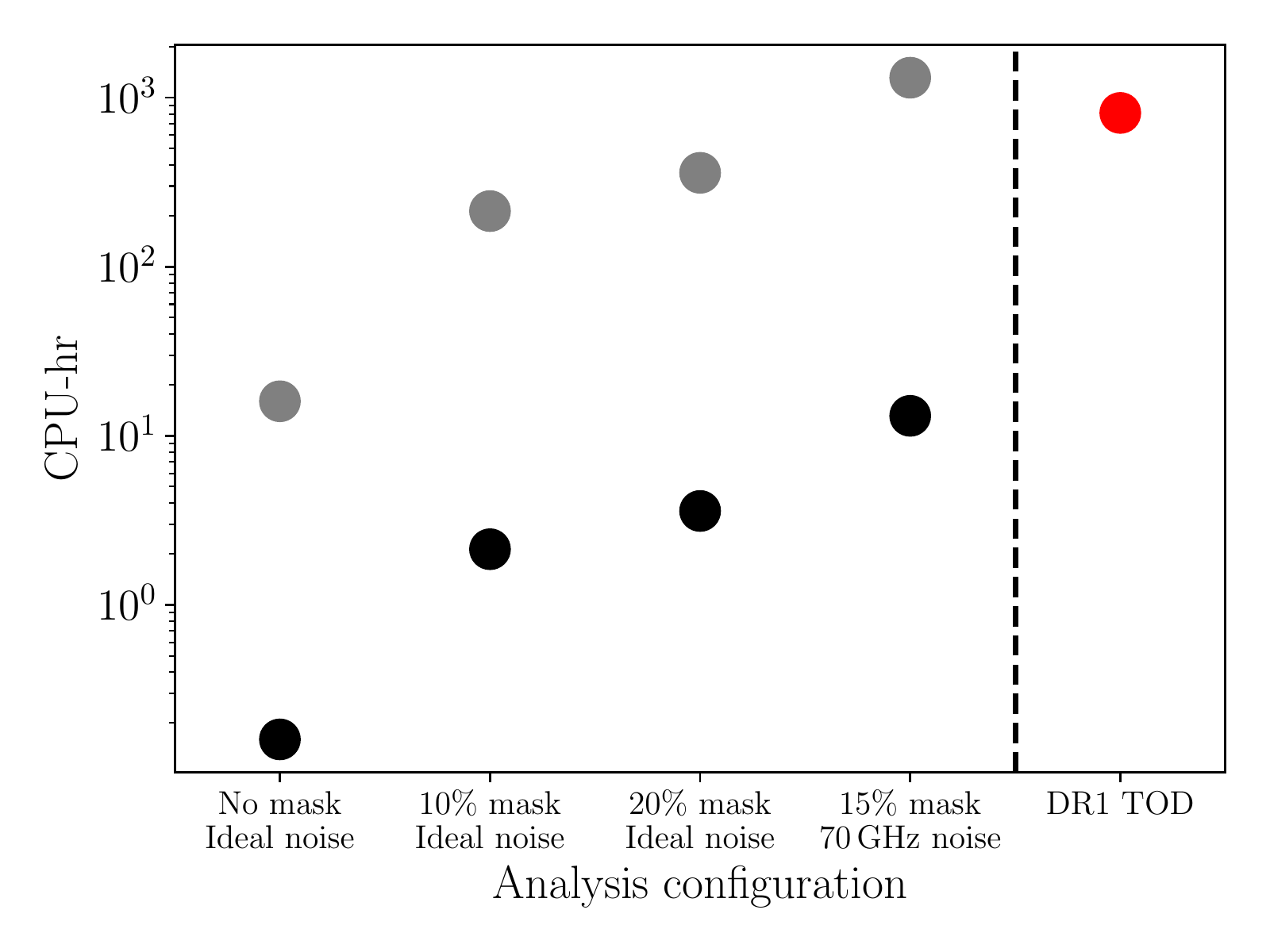}
	\caption{\label{fig:runtime}CPU-hours per accepted sample. The black dots are the costs per sample, while the gray dots are the costs per effective sample, i.e., normalized by the correlation length of the chain. This is compared to the 812 CPU-hours to generate one single end-to-end sample of DR1  \cosmoglobe\ \planck\ LFI and WMAP channels, in red.}
\end{figure}

The basic computational cost per sample for \commanderthree\ is plotted as black points in Fig.~\ref{fig:runtime}. Here we see that the overall cost increases rapidly with decreasing sky coverage, starting at 0.2\,CPU-hours for a full-sky sample to about 4\,CPU-hours for a 20\,\% constant latitude mask and uniform noise. As the isotropic noise cases have twice the noise of the mean of the 70\,GHz rms map, the higher signal-to-noise causes an increase in runtime, and for the realistic LFI 70\,GHz scanning pattern and 15\,\% mask, the cost is 20\,CPU-hours/sample. Taking into account that the overall correlation length is $\sim$\,100, the effective cost is therefore about 2\,000\,CPU-hours per independent sample; the gray dots in Fig.~\ref{fig:runtime} indicates this total cost for each configuration.

To interpret the implications of these costs, it is useful to compare with the overall cost of the full end-to-end processing pipeline within which this algorithm is designed to operate. The most relevant comparison point is therefore the cost of the \cosmoglobe\ DR1 processing, which is 812\,CPU-hours for the combination of \Planck\ LFI and WMAP, marked as a red dot in Fig.~\ref{fig:runtime}. In its current state, we find that the new cosmological parameter step alone is thus about 2.5 times more expensive than the full end-to-end processing.

\section{Summary and discussion}
\label{sec:conclusions}

This work had two main goals. The first goal was simply to implement the cosmological parameter sampling algorithm proposed by \citet{racine:2016} into the state-of-the-art \commanderthree\ CMB Gibbs sampler and demonstrate that this new step works in a production environment. A head-to-head comparison with both \cobaya\ and a stand-alone Python implementation demonstrates the fidelity of the method. We therefore conclude that this goal has been met, and as such this work finally adds true ``end-to-end'' processing abilities to \commanderthree. 

The second goal was to measure the computational performance of the method, assess whether it is already suitable for large-scale production, and identify potential bottlenecks that should be resolved in future work. In this case, we find that the overall cost per independent Gibbs sample for a data set similar to \cosmoglobe\ DR1 is about 2\,000\,CPU-hours, which is more than a factor of two higher than the total cost of all other steps in the end-to-end algorithm, including low-level calibration, map-making, and component separation. Therefore, although the code is technically operational at the current time, we also conclude that it is still too expensive to be useful in a full-scale production environment.

At the same time, this analysis has also revealed the underlying origin of these high expenses, which can be divided into two parts. First, the overall cost per sample is dominated by the expense for solving for the mean field (Eq.~\eqref{eq:mean-field-map}) and fluctuation maps (Eq.~\eqref{eq:fluc-map}) using conjugate gradients. In this respect, we note that the current implementation uses a standard diagonal preconditioner to solve these equations, as described by \citet{eriksen:2004}. However, after more than two decades of development, far more efficient preconditioners have been explored and described in the literature, including multi-resolution or multi-grid methods \citep[e.g.,][]{seljebotn:2013,seljebotn:2019}. Implementing one of them can potentially reduce the basic cost per sample by one or two orders of magnitude. The second issue is a long correlation length. We note that the current algorithm is based on a very basic random-walk Metropolis sampler, which generally performs poorly for highly correlated parameters; in the cases considered here, the correlation between the amplitude of scalar perturbations, $A_s$, and the optical depth of reionization, $\tau$, is a typical example of such a correlation. This suggests that the correlation length can be greatly reduced in several ways, for instance either by adding new data sets (such as weak gravitational lensing constraints) or by implementing a more sophisticated sampling algorithm that uses derivative information, such as a Hamiltonian sampler. Overall, we consider it very likely that future work will be able to reduce the total cost by one or two orders of magnitude, as needed for production, and we hope that the current results can provide inspiration and motivation for experts in the field to join the effort.

Finally, before concluding this paper, we note that a key application of the Bayesian end-to-end analysis framework as pioneered by \BP\ and \cosmoglobe\ is the analysis of next-generation CMB $B$-mode experiments, for instance LiteBIRD. In this case, it is worth noting that the polarization-based signal-to-noise ratio of the tensor-to-scalar ratio, $r$, is much lower than \Planck\ and WMAP's temperature signal-to-noise ratio to the $\Lambda$CDM parameters, and both the conjugate gradient cost and Markov chain correlation length are likely to be much shorter than for the case considered in this paper. As such, it is conceivable that the current method already performs adequately for LiteBIRD, and this will be explored in future work.

\begin{acknowledgements}
  The current work has received funding from the European
  Union’s Horizon 2020 research and innovation programme under grant
  agreement numbers 819478 (ERC; \textsc{Cosmoglobe}) and 772253 (ERC;
  \textsc{bits2cosmology}).
  In
  addition, the collaboration acknowledges support from
  RCN (Norway; grant no.\ 274990). Moreover, Simone Paradiso aknowledges
  support from the Government of Canada's New Frontiers in Research Fund,
  through grant NFRFE-2021-00595.
  We acknowledge the use of the Legacy Archive for Microwave Background Data
  Analysis (LAMBDA), part of the High Energy Astrophysics Science Archive Center
  (HEASARC). HEASARC/LAMBDA is a service of the Astrophysics Science Division at
  the NASA Goddard Space Flight Center.  
  Some of the results in this paper have been derived using the \texttt{healpy}
  and \texttt{HEALPix}\footnote{\url{http://healpix.sf.net}} packages
  \citep{gorski2005, Zonca2019}.
\end{acknowledgements}

\bibliographystyle{aa}

\bibliography{./bib/Planck_bib,./bib/CG_bibliography}

\appendix

\section{Analytic expression for a constant latitude mask}
\label{sec:appendixA}

The goal of this appendix is to show how we can calculate $N_{\ell m \ell' m'}^{-1}$ analytically for a constant latitude mask and uniform noise. This expression makes the map-making equation, in Eq.~\eqref{eq:mapmakingeq}, computationally faster to solve as we derive our expression to be $N_{\ell m \ell' m'}^{-1} \propto \delta_{mm'}$. These results are generally known, but we re-derive them here for completeness.

For uniform instrumental noise with no mask, the pixel covariance matrix is $N_{pp'} = \sigma^2 \delta_{pp'}$ where $\sigma$ is the white noise uncertainty per pixel. Including the constant latitude mask, the inverse of the noise covariance matrix can be written in pixel space as
$$
\left(N^{-1} \right)_{pp'} = \frac{1}{\sigma^2} \delta_{pp'} H(|\theta(p) -\pi/2|-b).
$$
Here, $H$ is the Heaviside function, meaning that we mask every pixel $p$ where $|\theta(p) -\pi/2| < b$ for some latitude $b$ in radians. For a masked pixel, we have $\left(N^{-1} \right)_{pp}=0$, meaning that the noise at that pixel is infinite, $N_{pp} = \infty$.

Defining the spherical harmonics functions $Y_{\ell m}\left(p\right) = Y_{\ell m}\left(\hat{n}(p)\right)$ and $\hat{n}(p) = (\theta(p), \phi(p))$ are the spherical coordinates at pixel $p$, which we can transform into spherical harmonics space,
\begin{align}
\nonumber
\left(N^{-1}\right)_{\ell m \ell' m'} &= \sum_{p p'}\left(N^{-1}\right)_{pp'}Y^{*}_{\ell m}(p)Y_{\ell' m'}(p')\\
\nonumber
&= \frac{1}{\sigma^2}\sum_{p p'} Y^{*}_{\ell m}(p)Y_{\ell' m'}(p') \delta_{pp'} H(|\theta -\pi/2|-b)\\
\nonumber
&= \frac{1}{\sigma^2}\sum_{p } Y^{*}_{\ell m}(p)Y_{\ell' m'}(p) H(|\theta -\pi/2|-b)\\
&= \frac{1}{\sigma^2}\sum_p \Tilde{P}_{\ell m}\Tilde{P}_{\ell' m'} e^{-i(m-m')\phi} H(|\theta -\pi/2|-b).
\end{align}
Here, we have defined $\Tilde{P}_{\ell m}=\Tilde{P}_{\ell m}(\theta) = \Delta_{\ell m}P_{\ell m}(\cos(\theta))$, where ${\Delta_{\ell m}=(-1)^m \sqrt{\frac{2\ell+1}{4\pi}\frac{(\ell - m)!}{(\ell+m)!}}}$ and $P_{\ell m}(\cos(\theta))$ are the associated Legendre polynomials. Thus, we write the spherical harmonics functions as $Y_{\ell m}(p) = \Tilde{P}_{\ell m} e^{im\phi}$.

We now switch from discrete pixels to continuous space, meaning that we change the sum to an integration where we account for the number of pixels, $N_{\mathrm{pix}}$, per area element,
\begin{equation}
\sum_p \rightarrow \frac{N_{\mathrm{pix}}}{4\pi}\int d\Omega  = \frac{N_{\mathrm{pix}}}{4\pi}\int_{0}^{2\pi} d\phi \int_{0}^{\pi} d\theta \sin(\theta).
\end{equation}
This gives us
\begin{align}
\nonumber
\left(N^{-1}\right)_{\ell m \ell' m'} &= \frac{N_{\mathrm{pix}}}{4\pi \sigma^2}\int_{0}^{2\pi} d\phi \int_{0}^{\pi} d\theta \sin(\theta)\Tilde{P}_{\ell m}  \Tilde{P}_{\ell' m'}  e^{-i(m-m')\phi}
\\
\nonumber
&\cdot H(|\theta -\pi/2|-b)\\
\nonumber
&= \frac{N_{\mathrm{pix}}}{2\sigma^2} \delta_{mm'}\int_{0}^{\pi} d\theta \sin(\theta)\Tilde{P}_{\ell m}  \Tilde{P}_{\ell' m'} H(|\theta -\pi/2|-b)\\
\nonumber
&= \frac{N_{\mathrm{pix}}}{2\sigma^2} \delta_{mm'}\\
\label{eq:N_lml'm'_step}
&\cdot \left(\int_{0}^{\pi/2-b} d\theta \sin(\theta) \Tilde{P}_{\ell m}  \Tilde{P}_{\ell' m'}+\int_{\pi/2+b}^{\pi} d\theta \sin(\theta)\Tilde{P}_{\ell m}  \Tilde{P}_{\ell' m'}\right).
\end{align}
Writing $x=\cos(\theta)$, we know that the associated Legendre polynomials $P_{\ell m}(x)$ are either symmetric or antisymmetric in $x\rightarrow-x$. $P_{\ell m}(x)$ is symmetric in $x \rightarrow -x$ when $\ell+m$ = even and antisymmetric when $\ell+m$ = odd. Since $m=m'$, we note that the two integrals in the last line of Eq.~\eqref{eq:N_lml'm'_step} cancel each other if $\ell+\ell' =$ odd. We, therefore, only get nonzero elements when $\ell + \ell' =$ even, for which the two integrals are equal. Hence, for $\ell + \ell'=$ even, we get
\begin{align}
\nonumber
\left(N^{-1}\right)_{\ell m \ell' m'} &= \frac{N_{\mathrm{pix}}}{\sigma^2} \delta_{mm'}\int_{0}^{\pi/2-b} d\theta \sin(\theta)\Tilde{P}_{\ell m}(\theta)\Tilde{P}_{\ell' m'}(\theta)\\
\label{eq:finished_n_inv}
&=\frac{N_{\mathrm{pix}}}{\sigma^2} \delta_{mm'}\int_{\sin(b)}^{1} dx \, \Tilde{P}_{\ell m}(\arccos(x)) \Tilde{P}_{\ell' m}(\arccos(x)).
\end{align}
In the full-sky case, $b=0$, we find $\left(N^{-1}\right)_{\ell m \ell' m'} = \delta_{\ell \ell'}\delta_{mm'}\frac{N_{\mathrm{pix}}}{4\pi \sigma^2}$ which is equivalent to a diagonal noise power spectrum ${N_{\ell} = \frac{4\pi\sigma^2}{N_{\mathrm{pix}}}}$.

The integral in Eq.~\eqref{eq:finished_n_inv} can be solved numerically by gridding $x$, and $\Tilde{P}_{\ell m}$ can be calculated in Python using the library \textit{pyshtools} \citep{https://doi.org/10.1029/2018GC007529}. The azimuthally symmetric Galactic mask used for this work has a sky coverage of $f_{\mathrm{sky}} = 0.90$, giving $\sin(b) = 0.1$. Therefore, we require many fewer grid points for $x$ if we use the following identity for $\ell+\ell' = $ even,
\begin{align}
&\delta_{mm'} \int_{\sin(b)}^{1} dx \, \Tilde{P}_{\ell m}(\arccos(x)) \Tilde{P}_{\ell' m}(\arccos(x)) = \\
\frac{1}{4\pi}\delta_{\ell \ell'}\delta_{m m'} - &\delta_{mm'} \int_{0}^{\sin(b)} dx \, \Tilde{P}_{\ell m}(\arccos(x)) \Tilde{P}_{\ell' m}(\arccos(x)),
\end{align}
which comes from the orthonormality condition for spherical harmonics. We now only need to grid $x$ in the interval $0\leq x \leq \sin(b) = 0.1$, as opposed to the interval $0.1 \leq x \leq 1$.

Since $\N^{-1}_{\ell m \ell' m'} \propto \delta_{m m'}$, we get simplified matrix expressions. Imagine multiplying the matrix $\N^{-1} = N^{-1}_{\ell m \ell' m'}$ by the vector $\vec{b} = b_{\ell m}$:
\begin{align}
\N^{-1} \cdot \vec{b} &= \sum_{\ell' m'}\left(\N^{-1}\right)_{\ell m \ell' m'}b_{\ell' m'} = \sum_{\ell'}\left(N^{-1}\right)_{\ell m \ell' m}b_{\ell' m}\\
&= \sum_{\ell' }\left(\N^{-1}\right)^{(m)}_{\ell, \ell'}b^{(m)}_{\ell'}.
\end{align}
To solve the map-making equation, Eq.~\eqref{eq:mapmakingeq}, we get a matrix equation for each $0 \leq m \leq \ell_{\mathrm{max}}$. This gives us $\ell_{\textrm{max}}+1$ number of matrix equations where the dimensions of the matrices are maximally $(\ell_{\textrm{max}}+1) \times (\ell_{\textrm{max}}+1)$. This is numerically much quicker to invert rather than inverting the full matrix $\left[\S^{-1} + \B^T \N^{-1}\B \right]$ once.

\end{document}